\documentclass[reqno]{amsart}

\usepackage{amsmath}
\usepackage[T1]{fontenc}
\usepackage{lmodern} 
\usepackage[utf8]{inputenc}
\usepackage{color}
\usepackage{graphicx}

\begin{document}

\title{Analytic solution of Guyer-Krumhansl equation for laser flash experiments}

\author{R. Kovács$^{123}$}

\address{$^1$Department of Theoretical Physics, Wigner Research Centre for Physics,
Institute for Particle and Nuclear Physics, Budapest, Hungary and \\
$^2$Department of Energy Engineering, Faculty of Mechanical Engineering, BME, Budapest, Hungary and \\
$^3$Montavid Thermodynamic Research Group}
\date{\today}

\begin{abstract}
The existence of non-Fourier heat conduction is known for a long time in small and low temperature systems. The deviation from Fourier's law has been found at room temperature in heterogeneous materials like rocks and metal foams \cite{Botetal16, Vanetal17}. These experiments emphasized that the so-called Guyer-Krumhansl equation is adequate for modeling complex materials. In this paper an analytic solution of Guyer-Krumhansl equation is presented considering boundary conditions from laser flash experiment. The solutions are validated with the help of a numerical code \cite{KovVan15} developed for generalized heat equations.

\end{abstract}
\maketitle


\section{Introduction}
\label{intro}

The existence of non-Fourier heat conduction under various conditions is experimentally proved in several different ways. First, the Maxwell-Cattaneo-Vernotte equation (MCV) \cite{Max1867, Cattaneo58, Vernotte58},
\begin{equation}
\tau_q \partial_{tt} T + \partial_t T = \alpha \partial_{xx} T,
\label{MCV}
\end{equation}
is used to describe the dissipative wave form of heat propagation called second sound. Here, $\tau_q$ is the relaxation time, $\alpha$ stands for the thermal diffusivity, $\partial_t$ denotes the time derivative and $\partial_{xx}$ denotes the second spatial derivative in one dimension. It is the simplest extension of Fourier's law and there are several different theorems in the literature which lead to this type of hyperbolic generalization \cite{JosPre89, JosPre90a, Gyar77a, JouVasLeb88ext, Tzou95, MulRug98, VanFul12, KovVan15, BerVan15, Cimmelli09nl, Cimm09diff}. The existence of second sound was predicted by Tisza and Landau \cite{Tisza38, Lan47}, earlier than the experimental discovery. Then Peshkov managed to measure it in superfluid He \cite{Pesh44} and enhanced the researches in that respect. Later on, several new ideas have developed how to measure similar phonemena in different materials. One of the most important result is related to Guyer and Krumhansl who derived the so-called window condition, significantly supporting the measurement of second sound in solids \cite{GK64}.

The next extension of Fourier's equation bears their names, called Guyer-Krumhansl (GK) equation \cite{GuyKru66a1, GuyKru66a2, Van01a},
\begin{equation}
\tau_q \partial_{tt} T + \partial_t T = \alpha \partial_{xx} T + \kappa^2 \partial_{txx} T,
\label{GK}
\end{equation}
where $\kappa^2$ is the dissipation parameter \cite{KovVan15}, strongly related to the mean free path from the aspect of kinetic theory \cite{MulRug98}. It contains the MCV equation (\ref{MCV}), however, it is a parabolic type model and is able to recover the solution of Fourier equation when $\kappa^2 / \tau = \alpha$ holds, called Fourier resonance \cite{Botetal16, Vanetal17, VanKovFul15}. Despite of the disadvantageous infinite propagation speed of parabolic models, it is still a valid and thermodynamically consistent realisation of non-Fourier heat conduction at room temperature \cite{Botetal16, Vanetal17, KovVan18dpl}.  

Regarding the experiments, one should mention the ballistic-type heat conduction measured by Jackson et al. \cite{JacWalMcN70, JacWal71, McNEta70a, McN74t} in NaF crystals and modeled by several authors \cite{DreStr93a, Ma13a, Ma13a1, Ma13a2}. The most recent one can be found in \cite{KovVan18} where quantitative agreement is obtained between the theory and experiments. The theory is based on non-equilibrium thermodynamics with internal variables and Nyíri multipliers \cite{KovVan15, BerVan15, Nyiri91}. 

The experimental success of measuring the second sound and the universal theory of non-equilibrium thermodynamics has motivated the researchers to find non-Fourier heat conduction in wave form described by the MCV equation (\ref{MCV}) at room temperature. For example, such an endeavor is related to the experiments of Mitra et al. \cite{MitEta95} where a frozen meat is used to find similar phenomenon. Unfortunately, no one was able to reproduce these experimental results and the measurements of Mitra et al. are widely criticized \cite{TilVic09, HerBec00, HerBec00b}. However, it turned out that the GK equation could be the relevant measurable extension of Fourier's law, the related non-Fourier effects are measured several times in different materials \cite{Botetal16, Vanetal17}. In many other cases the dual phase lag model is considered also as an adequate generalization \cite{TanEtal07, AkbPas14, AfrinEtal12, LiuChen10, Zhang09}, however, this model is contradictory to basic physical principles \cite{KovVan18dpl} and its validity is questionable \cite{Ruk14, Ruk17, Fabetal14, FabLaz14a, FabEtal16, Quin07, ChirCiaTib17}. 

All the aforementioned experiments are the heat pulse type, the underlying principle is the same, only the equipment is different. It is a standard method to measure the thermal diffusivity and is used widely in engineering practice. 
The importance of Guyer-Krumhansl equation (\ref{GK}) in the evaluation of such experiments indicated the need to find an analytic solution. 

The work of Zhukovsky has to be mentioned here \cite{Zhukov16, Zhu16a, Zhu16b, ZhuSri17}. Recently, Zhukovsky obtained an exact solution of GK equation using operational method for infinite spatial domain. Moreover, different initial conditions are considered, the wave-like initial condition together with decaying boundary conditions have greater importance. Despite of these valuable results, it is still quite far from the experiments. 
Therefore, the goal of this paper is to complement the results of the aforementioned papers to be more applicable for real experimental setup like described below.

\section{Experimental setup and boundary conditions}
\label{bcs}

\begin{figure}[h]
\centering
\includegraphics[width=10cm,height=8cm]{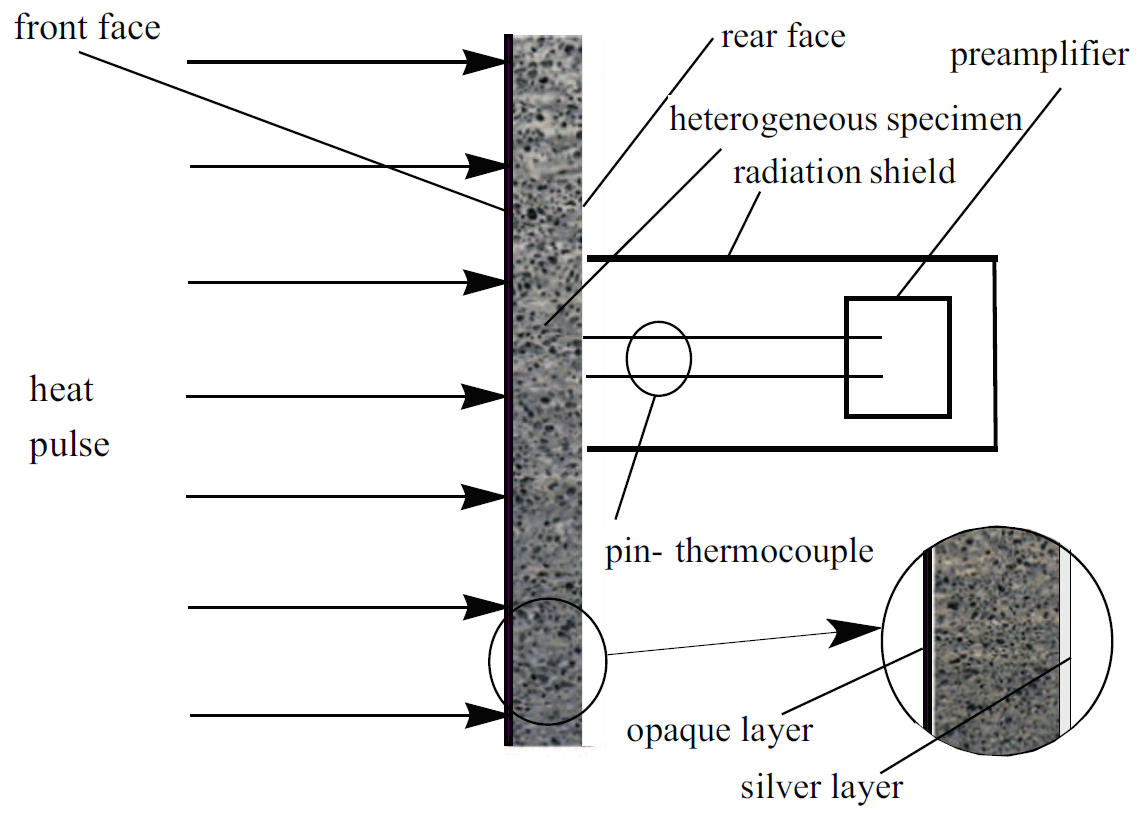}
\caption{Arrangement of  the experiment, original figure from \cite{Vanetal17}.}
\label{fig:exp1}
\end{figure}

Measurements finding non-Fourier heat conduction in heterogeneous materials are performed on room temperature as it is described in detail in the papers \cite{Botetal16, Vanetal17} have the following setup, see Fig. \ref{fig:exp1}. 

The front side boundary condition depicts the heat pulse which excites the heterogeneous sample. The pulse has a finite length, given as $t_p=0.01$ s \cite{Botetal16, Vanetal17}. The exact shape of the pulse has not been taken in account in \cite{Botetal16, Vanetal17} during the evaluation process, nevertheless, its length is critical and greatly influences the solution \cite{GrofPhD02}. As it is highlighted and applied in \cite{Botetal16, Vanetal17, KovVan15, BCTFGGPV13}, the following function is considered to model the heat pulse,
\begin{center}
	$q( x=0, t)= \left\{ \begin{array}{cc}
	q_{max} \left(1-cos\left(2 \pi \cdot \frac{ t}{t_p}\right)\right) &
	\textrm{if } 0<  t \leq  t_p,\\
	0 & \textrm{if }  t>  t_p,
            \end{array} \right.  $
\end{center}
that is, the front side boundary condition is given by prescribing the heat flux in time, here $q_{max}$ is the amplitude of the signal. When the experimental results are evaluated, the cooling on boundary had to be considered. Nevertheless it is crucial to model these effects, in the analytic solution it is neglected to simplify the mathematical problem. Thereby adiabatic condition is applied to the rear side for every time instant $q(x=L, t) = 0$. 
Regarding the initial conditions, all the time derivatives are zero at the initial state and the sample is in equilibrium with its environment, i.e. $T(x,t=0) = T_0$. 

\section{Dimensionless quantities}

In order to ease the solution of GK equation dimensionless quantities are used (see \cite{KovVan15} for details). From now on, the same formalism is applied, that is, the following parameters are introduced,
\begin{eqnarray}
\hat{t} =\frac{\alpha t}{L^2} \quad &\text{with}& \quad
\alpha=\frac{\lambda}{\rho c};  \quad
\hat{x}=\frac{x}{L};\nonumber \\
\hat{T}=\frac{T-T_{0}}{T_{\text{end}}-T_{0}} \quad &\text{with}&\quad
T_{\text{end}}=T_{0}+\frac{\bar{q}_0 t_p}{\rho c L};  \nonumber \\
\hat{q}=\frac{q}{\bar{q}_0} \quad &\text{with}&\quad
\bar{q}_0=\frac{1}{t_p}  \int_{0}^{t_p} q_{0}(t)dt,
\label{ndvar}\end{eqnarray}
where $L$ is the length of the sample, $\lambda$, $\rho$ and $c$ are the thermal conductivity, mass density and specific heat, respectively. The time averaged heat flux $\bar q_0$ is used to define the equilibrium temperature $T_{end}$. The material parameters converted with
\begin{equation}
\hat{\tau}_\Delta =\frac{\alpha t_p}{L^2}; \quad
\hat{\tau}_q 	  = \frac{\alpha \tau_{q}}{L^2}; \quad
\hat{\kappa} 	  = \frac{\kappa}{L},
\end{equation}
where $\hat{\tau}_\Delta$ stands for the dimensionless heat pulse length and $\hat \tau_q$ denotes the relaxation time related to the heat flux. For the sake of simplicity, the notation ``hat'' is omitted and let us restrict ourselves only for dimensionless quantities. 
Using these formalism the GK-type heat equation reads as
\begin{equation}
\tau_q \partial_{tt} T + \partial_t T = \partial_{xx} T + \kappa^2 \partial_{txx} T,
\label{ndGK}
\end{equation}
which can be decomposed into two equations containing the balance equation of internal energy
\begin{equation}
\tau_{\Delta} \partial_t T + \partial_x q = 0,
\label{ndbalen}
\end{equation}
and the GK-type consititutive equation is:
\begin{equation}
\tau_q \partial_t q + q +\tau_{\Delta} \partial_x T - \kappa^2 \partial_{xx} q=0.
\label{ndcongkeq}
\end{equation}
Since the boundary conditions are prescribed as a given heat flux in time it is suitable to eliminate $T$ from the equations (\ref{ndbalen}) and (\ref{ndcongkeq}):
\begin{equation}
\tau_q \partial_{tt} q + \partial_t q = \partial_{xx} q + \kappa^2 \partial_{txx} q.
\label{ndGKforq}
\end{equation}
After obtaining the solution for $q(x,t)$ one can use eq. (\ref{ndbalen}) to integrate $\partial_x q$ respect to time and calculate $T(x,t)$. Applying dimensionless quantities, the heat pulse boundary condition at the front side reads as
\begin{center}
	$q( x=0, t)=q_0(t)= \left\{ \begin{array}{cc}
	\left(1-cos\left(2 \pi \cdot \frac{ t}{\tau_{\Delta}}\right)\right) &
	\textrm{if } 0<  t \leq  \tau_{\Delta},\\
	0 & \textrm{if }  t>  \tau_{\Delta},
            \end{array} \right.  $
\end{center}
and for the rear side $q(x=1,t)=q_L(t)=0$ holds together with the dimensionless initial condition $T(x,t=0)=0$. 

\section{Solution method}
According to the front side boundary condition it is reasonable to split the solution into two sections in time. The first one goes from $0$ to $\tau_{\Delta}$ and the second interval starts at $\tau_{\Delta}$ and reaches up to an arbitrary time instant $t$. The basic mathematical principles and procedures can be found in \cite{CarJae59b, Farlow93b, GTvN11b}. The sample length $L$ is intentionally left unchanged in the following as it highlights the integration limits in non-dimensionless formalism. In case of dimensionless quantities $L$ can be simply considered as $L=1$, since $0\leq x \leq 1$.

\subsection{Section I. ($0<t<\tau_{\Delta}$)}
Due to the time dependent boundary condition, let us split the solution of $q(x,t)$ as
\begin{equation}
q(x,t) = w(x,t) + v(x,t),
\label{qbont}
\end{equation}
where $w(x,t)$ is used to separate the time dependence of the boundary condition from the part $v(x,t)$. It is arbitrary to choose the form of $w(x,t)$, for the sake of simplicity it is satisfactory to assume its form to be linear, i.e.
\begin{equation}
w(x,t) := q_0(t) + \frac{x}{L} \big ( q_L(t) - q_0(t) \big ) = \big (1-\frac{x}{L} \big ) q_0(t),
\end{equation}
as the rear side is adiabatic. For further calculations let us simplify and shorten our notation of partial derivatives: $\partial_t = \dot \Box$ and $\partial_x = \Box ' $.
Substituting (\ref{qbont}) into (\ref{ndGKforq}), it yields
\begin{equation}
\tau_q (\ddot w + \ddot v) + \dot w + \dot v = w'' + v'' + \kappa^2 (\dot w'' + \dot v'').
\label{eq1}
\end{equation}
Therefore $v(x,t)$ has constant boundary condition in time but an inhomogeneous term appears since  $\dot w =  \big (1-\frac{x}{L} \big ) \dot q_0(t)$, $\ddot w =  \big (1-\frac{x}{L} \big ) \ddot q_0(t)$ holds and $w'' = 0$.
At this point one obtains an inhomogeneous equation for $v(x,t)$,
\begin{equation}
\tau_q  \ddot v + \dot v =  v'' + \kappa^2 \dot v'' - f(x,t),
\label{eq2}
\end{equation}
where $f(x,t) = \dot w + \tau_q \ddot w$. The splitting (\ref{qbont}) preserves the initial conditions: $v(x,t=0) = 0$ and $\dot v(x,t=0)=0$. Regarding the boundary conditions, $v(x=0,t)=0$, $v(x=L,t)=0$ holds. 
The inhomogeneous term $f(x,t)$ can be determined from $q_0(t)$ as
\begin{eqnarray}
\dot w = \frac{2 \pi}{\tau_{\Delta}} \big (1-\frac{x}{L} \big ) \sin (2 \pi \frac{t}{\tau_{\Delta}}), \\
\ddot w = \frac{4 \pi^2 }{\tau_{\Delta}^2} \big (1-\frac{x}{L} \big ) \cos (2 \pi \frac{t}{\tau_{\Delta}}).
\end{eqnarray}
Let us suppose now that the variables can be separated and
\begin{equation}
v(x,t) = \varphi (t) X(x)
\label{vbont}
\end{equation}
exists and dissociate the partial differential equation (\ref{eq2}) into two ordinary differential equations (ODEs). As equation (\ref{eq2}) is inhomogeneous one should also assume that the eigenfunctions $X(x)$ of the homogeneous case ($f(x,t)=0$) solves the inhomogeneous equation, too. This system of eigenfunctions is used to explicate $f(x,t)$ in the function space spanned by the solutions $X(x)$. Thus, one has to calculate the homogeneous part of $v(x,t)$ that is done as follows. The separation of variables, eq. (\ref{vbont}), leads to the equation for homogeneous part
\begin{equation}
\frac{\tau_q \ddot \varphi + \dot \varphi}{\varphi + \kappa^2 \dot \varphi} = \frac{X''(x)}{X(x)} = - \beta, \quad \beta \in \mathbb{R}^+,
\end{equation}
hence the eigenfunctions are determined by the equation
\begin{equation}
X'' + \beta X =0, \quad X(x=0)=0, \quad X(x=L) =0.
\end{equation}
The general solution reads as 
\begin{equation}
X(x)=A \cos (\sqrt{\beta }x ) + B \sin (\sqrt{\beta }x ),
\end{equation}
where the constants $A$ and $B$ are determined according to the boundary conditions for $v(x,t)$. The condition  $X(x=0)=0$ implies that $A=0$ and $X(x=L)$ determines the eigenvalues. As $B\neq0$, otherwise it would lead to a trivial solution, $\sin(\sqrt{\beta }x )=0$ holds, hence 
\begin{equation}
\beta_n = \big ( \frac{n \pi}{L} \big )^2,
\end{equation}
where $0<n \in \mathbb{N}$. In summary, 
\begin{equation}
X_n(x) = \sin \big ( \frac{n \pi}{L} x \big )
\label{sfv}
\end{equation}
is an eigenfunction of the operator $\frac{d^2}{dx^2}$ with positive eigenvalues $\beta_n$. The constant $B$ will be combined with the  emerging solution of the time evolution part $\varphi (t)$. Using eq. (\ref{sfv}) one obtains 
\begin{equation}
v(x,t) = \sum\limits_{n=1}^{\infty} \varphi_n(t) \sin \big ( \frac{n \pi}{L} x \big ),
\end{equation}
that is, the inhomogeneous term $f(x,t)$ has to be accounted now,
\begin{eqnarray}
-f(x,t) &=& \tau_q \ddot v + \dot v - v'' -\kappa^2 \dot v'' = \\
&=& \sum\limits_{n=1}^{\infty} \big [\tau_q \ddot \varphi_n + \dot \varphi_n + \big ( \frac{n \pi}{L} \big )^2 \varphi_n + \kappa^2 \big ( \frac{n \pi}{L} \big )^2 \dot \varphi_n \big ] \sin \big ( \frac{n \pi}{L} x \big ). \label{eqvart}
\end{eqnarray}
It can be solved for every $n$ if the function $f(x,t)$ is decomposed according to the eigenfunctions, it yields an ODE for $\varphi_n$. The Fourier series of $f(x,t)$ is given as 
\begin{equation}
f(x,t) = \sum\limits_{n=1}^{\infty} f_n(t)  \sin \big ( \frac{n \pi}{L} x \big ),
\end{equation}
where 
\begin{equation}
f_n(t) = \frac{2}{L} \big [ \frac{2 \pi}{\tau_{\Delta}} \sin \big (2 \pi \frac{t}{\tau_{\Delta}} \big ) + \tau_q \frac{4 \pi^2 }{\tau_{\Delta}^2}  \cos \big (2 \pi \frac{t}{\tau_{\Delta}} \big) \big] \int\displaylimits_0^L \big (1-\frac{x}{L} \big ) \sin \big ( \frac{n \pi}{L} x \big ) dx.
\end{equation}
Calculating the integral on the right hand side yields
\begin{equation}
f_n(t) = \big [ \frac{2 \pi}{\tau_{\Delta}} \sin \big (2 \pi \frac{t}{\tau_{\Delta}} \big ) + \tau_q \frac{4 \pi^2 }{\tau_{\Delta}^2}  \cos \big (2 \pi \frac{t}{\tau_{\Delta}} \big) \big] \frac{2}{n \pi} = f(t) \frac{2}{n \pi},
\end{equation}
\begin{equation}
f(x,t) = \sum\limits_{n=1}^{\infty}f(t) \frac{2}{n \pi}  \sin \big ( \frac{n \pi}{L} x \big ).
\end{equation}
Now the resulted ODE can be solved for $\varphi_n(t)$ with initial conditions $\varphi_n(0)=0$ and $\dot \varphi_n(0) =0$:
\begin{equation}
\tau_q \ddot \varphi_n + \big ( 1 + \kappa^2 \big (\frac{n \pi}{L} \big )^2 \big ) \dot \varphi_n + \big (\frac{n \pi}{L} \big )^2  \varphi_n = -f(t)\frac{2}{n \pi}.
\end{equation}
Its solution is calculated using Wolfram Mathematica, it reads as
\begin{eqnarray}
\varphi_n(t) = \frac{1}{2 \sqrt{a^2-4 b} \left(a^2 g^2+\left(b-g^2\right)^2\right)}e^{-\frac{1}{2} \left(a+\sqrt{a^2-4 b}\right) t} \cdot \left(a^2 c \left(-1+e^{\sqrt{a^2-4 b} t}\right) g -\right. \nonumber \\
-\left(\sqrt{a^2-4 b} d \left(1+e^{\sqrt{a^2-4 b} t}\right)+2 c \left(-1+e^{\sqrt{a^2-4 b} t}\right) g\right) \left(b-g^2\right)+ \nonumber \\
+a \left(\sqrt{a^2-4 b} c g+\sqrt{a^2-4 b} c e^{\sqrt{a^2-4 b} t}g+d \left(b+g^2\right)-d e^{\sqrt{a^2-4 b} t} \left(b+g^2\right)\right)+ \nonumber \\
\left.+2 \sqrt{a^2-4 b} e^{\frac{1}{2} \left(a+\sqrt{a^2-4 b}\right) t} ((b d-g (a c+d g)) \cos(g t)+(b c+g (a d-c g)) \sin(g t))\right),
\end{eqnarray}
where the constants $a,b,c,d,g$ are given as
\begin{eqnarray}
a = \frac{1}{\tau_q} \big ( 1 + \kappa^2 \big (\frac{n \pi}{L} \big )^2 \big ), \quad b= \frac{1}{\tau_q} \big (\frac{n \pi}{L} \big )^2,  \nonumber \\
c =-\frac{4}{ n \tau_{\Delta} \tau_q}, \quad d = -\frac{8 \pi}{n \tau_{\Delta}^2}, \quad g = \frac{2 \pi}{\tau_{\Delta}}. 
\end{eqnarray}
Now $v(x,t)$ is obtained together with the solution of first section $q_I(x,t) = w(x,t) + v(x,t)$.
\newpage

\subsection{Section II. ($\tau_\Delta <t$)}
For section II, the initial condition is determined based on the functions $q_I(x,t=\tau_\Delta)$ and $\dot q_I(x,t=\tau_\Delta)$.
Here we seek for the solution of eq. (\ref{ndGKforq}) with time independent boundary conditions. These are prescribed as adiabatic condition on both sides. However, the initial conditions are more difficult to consider. Let us introduce $\tilde t $ as $\tilde t = t - \tau_\Delta$ to ease the calculations. The initial conditions are
\begin{equation}
q_{II}(x, \tilde t=0)=q_I(x,t=\tau_{\Delta}), \quad \dot q_{II}(x, \tilde t =0) = \dot q_I(x,t=\tau_{\Delta}).
\end{equation}
Moreover, the inhomogeneous term $f(x,t)$ is vanished for that section due to constant boundary conditions. Let us separate  the variables again and assume that 
\begin{equation}
q_{II}(x,\tilde t) = \gamma(\tilde t) X(x),
\end{equation}
where the eigenfunctions $X(x)$ and eigenvalues $\beta_n$ are already calculated in the previous section. In order to determine $\gamma(\tilde t)$ an ODE has to be solved,
\begin{equation}
\tau_q \ddot \gamma_n + (1 + \beta_n \kappa^2) \dot \gamma_n + \beta_n \gamma_n = 0
\label{eq21}
\end{equation}
with initial conditions $\gamma_n (0) = \varphi_n (\tau_\Delta)$ and $\dot \gamma_n (0) = \dot \varphi_n (\tau_\Delta)$.
Its general solution is
\begin{equation}
\gamma_n(\tilde t) = C_{1n} e^{r_{1n} \tilde t} + C_{2n} e^{r_{2n} \tilde t},
\end{equation}
where the characteristic exponents are
\begin{equation}
r_{1,2} = \frac{1}{2 \tau_q} \big ( -1 -\beta_n \kappa^2 \pm \sqrt{(1+\beta_n \kappa^2)^2 - 4 \tau_q \beta_n} \big ).
\end{equation}
Taking into account the initial conditions for the constants $C_{1n}$ and $ C_{2n}$, leads to
\begin{eqnarray}
C_{1n} + C_{2n} = \varphi_n (t=\tau_\Delta), \nonumber \\
C_{1n} r_{1n} + C_{2n} r_{2n}= \dot \varphi_n (t=\tau_\Delta).
\end{eqnarray}
It is solved again using Wolfram Mathematica where the $R = \sqrt{a^2 - 4b}$ notation is applied.

\begin{align}
&C_{1n} = -\frac{1}{r_1-r_2}\left(-\frac{1}{4 \left(a^2 g^2+\left(b-g^2\right)^2\right) R}e^{-\frac{1}{2} (a+R) \tau_\Delta} (-a-R)\cdot\right. \nonumber \\
&\cdot \left(a^2 c \left(-1+e^{R \tau_\Delta}\right) g+2 e^{\frac{1}{2} (a+R) \tau_\Delta} (b d-g (a c+d g)) R-\left(b-g^2\right) \left(2 c \left(-1+e^{R \tau_\Delta}\right) g+\right.\right. \nonumber \\
&+\left.\left.+d \left(1+e^{R \tau_\Delta}\right) R\right)+a \left(d \left(b+g^2\right)-d e^{R \tau_\Delta} \left(b+g^2\right)+c g R+c e^{R \tau_\Delta} g R\right)\right)- \nonumber \\
&-\frac{1}{2 \left(a^2 g^2+\left(b-g^2\right)^2\right) R}e^{-\frac{1}{2} (a+R) \tau_\Delta} \left(a^2 c e^{R \tau_\Delta} g R+2 e^{\frac{1}{2} (a+R) \tau_\Delta} g (b c+g (a d-c g)) R+\right. \nonumber \\
&+e^{\frac{1}{2} (a+R) \tau_\Delta} (b d-g (a c+d g)) R (a+R)-\left(b-g^2\right) \left(2 c e^{R \tau_\Delta} g R+d e^{R \tau_\Delta} R^2\right)+ \nonumber \\
&\left.+a \left(-d e^{R \tau_\Delta} \left(b+g^2\right) R+c e^{R \tau_\Delta} g R^2\right)\right)+\frac{1}{2 \left(a^2 g^2+\left(b-g^2\right)^2\right) R}e^{-\frac{1}{2} (a+R) \tau_\Delta} \cdot \nonumber \\
&\cdot \left(a^2 c \left(-1+e^{R \tau_\Delta}\right) g+2 e^{\frac{1}{2} (a+R) \tau_\Delta} (b d-g (a c+d g)) R-\right.\left(b-g^2\right) \left(2 c \left(-1+e^{R \tau_\Delta}\right) g+ \right.\nonumber \\
&+\left.\left.\left.d \left(1+e^{R \tau_\Delta}\right) R\right)+a \left(d \left(b+g^2\right)-d e^{R \tau_\Delta} \left(b+g^2\right)+c g R+c e^{R \tau_\Delta} g R\right)\right) r_2\right), \nonumber \\
&C_{2n} =\frac{1}{2 \left(a^2 g^2+\left(b-g^2\right)^2\right) R}e^{-\frac{1}{2} (a+R)\tau_\Delta} \left(a^2 c \left(-1+e^{R \tau_\Delta}\right) g+2 e^{\frac{1}{2} (a+R) \tau_\Delta} \cdot \right. \nonumber \\
&\cdot (b d-g (a c+d g)) R-\left(b-g^2\right) \left(2 c \left(-1+e^{R \tau_\Delta}\right) g+d \left(1+e^{R \tau_\Delta}\right) R\right)+a \left(d \left(b+g^2\right)-\right. \nonumber \\
&\left.\left.-d e^{R \tau_\Delta} \left(b+g^2\right)+c g R+c e^{R\tau_\Delta} g R\right)\right)+\frac{1}{r_1-r_2}\left(-\frac{1}{4 \left(a^2 g^2+\left(b-g^2\right)^2\right) R}\cdot \right. \nonumber \\
&\cdot e^{-\frac{1}{2} (a+R) \tau_\Delta} (-a-R) \left(a^2 c \left(-1+e^{R \tau_\Delta}\right) g+2 e^{\frac{1}{2} (a+R) \tau_\Delta} (b d-g (a c+d g)) R-\right. \nonumber \\
&-\left(b-g^2\right) \left(2 c \left(-1+e^{R \tau_\Delta}\right) g+d \left(1+e^{R \tau_\Delta}\right) R\right)+a \left(d \left(b+g^2\right)-d e^{R \tau_\Delta} \left(b+g^2\right)+c g R+\right. \nonumber \\
&\left.\left.+c e^{R \tau_\Delta} g R\right)\right)-\frac{1}{2 \left(a^2 g^2+\left(b-g^2\right)^2\right) R}e^{-\frac{1}{2} (a+R) \tau_\Delta} \left(a^2 c e^{R \tau_\Delta} g R+\right. \nonumber \\
&+2 e^{\frac{1}{2} (a+R) \tau_\Delta} g (b c+g (a d-c g)) R+e^{\frac{1}{2} (a+R) \tau_\Delta} (b d-g (a c+d g)) R (a+R)- \nonumber \\
&\left.-\left(b-g^2\right) \left(2 c e^{R \tau_\Delta} g R+d e^{R \tau_\Delta} R^2\right)+a \left(-d e^{R \tau_\Delta} \left(b+g^2\right) R+c e^{R \tau_\Delta} g R^2\right)\right)+\nonumber \\
&+\frac{1}{2 \left(a^2 g^2+\left(b-g^2\right)^2\right) R}e^{-\frac{1}{2} (a+R) \tau_\Delta} \left(a^2 c \left(-1+e^{R \tau_\Delta}\right) g+\right. \nonumber \\
&+2 e^{\frac{1}{2} (a+R) \tau_\Delta} (b d-g (a c+d g)) R-\left(b-g^2\right) \left(2 c \left(-1+e^{R \tau_\Delta}\right) g+\right. \nonumber \\
&\left.\left.\left.+d \left(1+e^{R \tau_\Delta}\right) R\right)+a \left(d \left(b+g^2\right)-d e^{R \tau_\Delta} \left(b+g^2\right)+c g R+c e^{R \tau_\Delta} g R\right)\right) r_2\right).
\end{align}

\section{Temperature distribution}

So far we have seen the solution for the field of heat flux $q$. It uniquely determines the temperature field by using the balance equation of internal energy, eq. (\ref{ndbalen}). Again, one has to perform the calculations for both sections. 
\begin{align}
\tau_{\Delta} \dot T + q' = 0, \Rightarrow
\dot T = -\frac{1}{\tau_{\Delta}} q' = -\frac{1}{\tau_{\Delta}} \sum\limits_{n=1}^{\infty} \Gamma_n(t) \frac{n \pi}{L} \cos \big ( \frac{n \pi}{L} x \big ), \\
T = -\frac{1}{\tau_{\Delta}} \int\displaylimits_0^{t} \sum\limits_{n=1}^{\infty} \Gamma_n(\alpha) \frac{n \pi}{L} \cos \big ( \frac{n \pi}{L} x \big ) d\alpha,
\end{align}
where $\Gamma_n(t)$ could be $\varphi_n$ or $\gamma_n$ depending on which section is considered.
The initial condition for temperature in section I is $T_I(x,t=0)=0$, for section II is $T_{II}(x,\tilde t=0)=T_I(x,t=\tau_\Delta)$. For section I it reads as 
\begin{align} 
T_I (x,t) &= -\frac{1}{\tau_{\Delta}} \int\displaylimits_0^{t} (w'(\alpha) + v'(\alpha)) d\alpha =& \nonumber \\&= -\frac{1}{\tau_{\Delta}} \int\displaylimits_0^{t}  \left ( -\frac{1}{L} q_0(\alpha) + \sum\limits_{n=1}^{\infty} \varphi_n(\alpha) \frac{n \pi}{L} \cos \big ( \frac{n \pi}{L} x \big ) \right) d\alpha, \\
\int\displaylimits_0^{t} w'(\alpha) d\alpha &= \frac{1}{L} \big ( -t + \frac{t_p \sin (2 \pi t / t_p)}{2 \pi} \big ),
\end{align}
\begin{align} 
&\int\displaylimits_0^{t}  \sum\limits_{n=1}^{\infty} \varphi_n(\alpha) d\alpha = \sum\limits_{n=1}^{\infty} \Phi_n(t)= \frac{1}{g \left(a^2 g^2+\left(b-g^2\right)^2\right) (a-R) R (a+R)} \cdot \nonumber\\
&\cdot \left((b c+g (a d-c g)) (a-R) R (a+R)+g(a+R) \left(a^2 c g-a d \left(b+g^2\right)+\right.\right. \nonumber\\
&+ \left.a c g R-\left(b-g^2\right) (2 cg+d R)\right)-g (a-R) \left(a^2 c g-\left(b-g^2\right) (2 c g-d R)-\right. \nonumber\\
&- \left.a \left(d \left(b+g^2\right)+c g R\right)\right)-e^{-\frac{1}{2} (a+R) t} \left(e^{Rt} g (a+R) \cdot\right. \nonumber\\
&\cdot(a^2 c g - a d (b + g^2) + a c g R - (b - g^2) (2 c g + d R)) - \nonumber \\
&-g (a-R) \left(a^2 c g-\left(b-g^2\right) (2 c g-d R)-a \left(d \left(b+g^2\right)+c g R\right)\right)+\nonumber \\
&\left.\left.+e^{\frac{1}{2} (a+R) t} (a-R) R (a+R) ((b c+g (a d-c g)) \sin(g t)+(-b d+g (a c+d g)) \sin(gt))\right)\right),
\end{align}
It follows from $\varphi_n(t=0) = 0$ that 
\begin{equation}
\sum\limits_{n=1}^{\infty} \Phi_n(t=0)=0
\end{equation}
is true at time instant $t=0$. The initial condition for section I is automatically fulfilled. In case of section II the temperature distribution has to be fitted for $T_I(x,t=\tau_\Delta)$, i.e.
\begin{equation}
T_{II}= -\frac{1}{\tau_{\Delta}} \int\displaylimits_0^{\tilde t} \sum\limits_{n=1}^{\infty} \gamma_n(\alpha) \frac{n \pi}{L} \cos \big ( \frac{n \pi}{L} x \big ) d\alpha,
\end{equation}
\begin{equation}
\int\displaylimits_0^{\tilde t} \sum\limits_{n=1}^{\infty} \gamma_n(\alpha) d\alpha = \sum\limits_{n=1}^{\infty} \frac{C_{1n}}{r_{1n}} \big (e^{r_{1n} \tilde t} -1 \big )=\sum\limits_{n=1}^{\infty} \Omega_n (\tilde t).
\end{equation}
For $\Omega_n(\tilde t=0) = 0$ holds thus one has to exploit the integration constant and determine its value to fulfill the initial condition. Let us consider now the integration constant $K_{n}$ which is calculated as follows:
\begin{align}
&T_{II}(x,\tilde t=0)=T_I(x,t=\tau_{\Delta}) = -\frac{1}{\tau_{\Delta}} \left ( -\frac{\tau_{\Delta}}{L} + \sum\limits_{n=1}^{\infty} \Phi_n(t=\tau_{\Delta}) \frac{n \pi}{L} \cos (\frac{n \pi}{L} x) \right ) = \nonumber \\
&= -\frac{1}{\tau_{\Delta}} \left (\sum\limits_{n=1}^{\infty} \Omega_n(\tilde t=0) \frac{n \pi}{L} \cos (\frac{n \pi}{L} x) \right ) + \sum\limits_{n=1}^{\infty} K_{n} \frac{n \pi}{L} \cos (\frac{n \pi}{L} x) + \frac{1}{L},
\end{align}
that is, $K_{n} = \Phi_n(t=\tau_\Delta)$.

\begin{figure}[h]
\includegraphics[width=12cm,height=7cm]{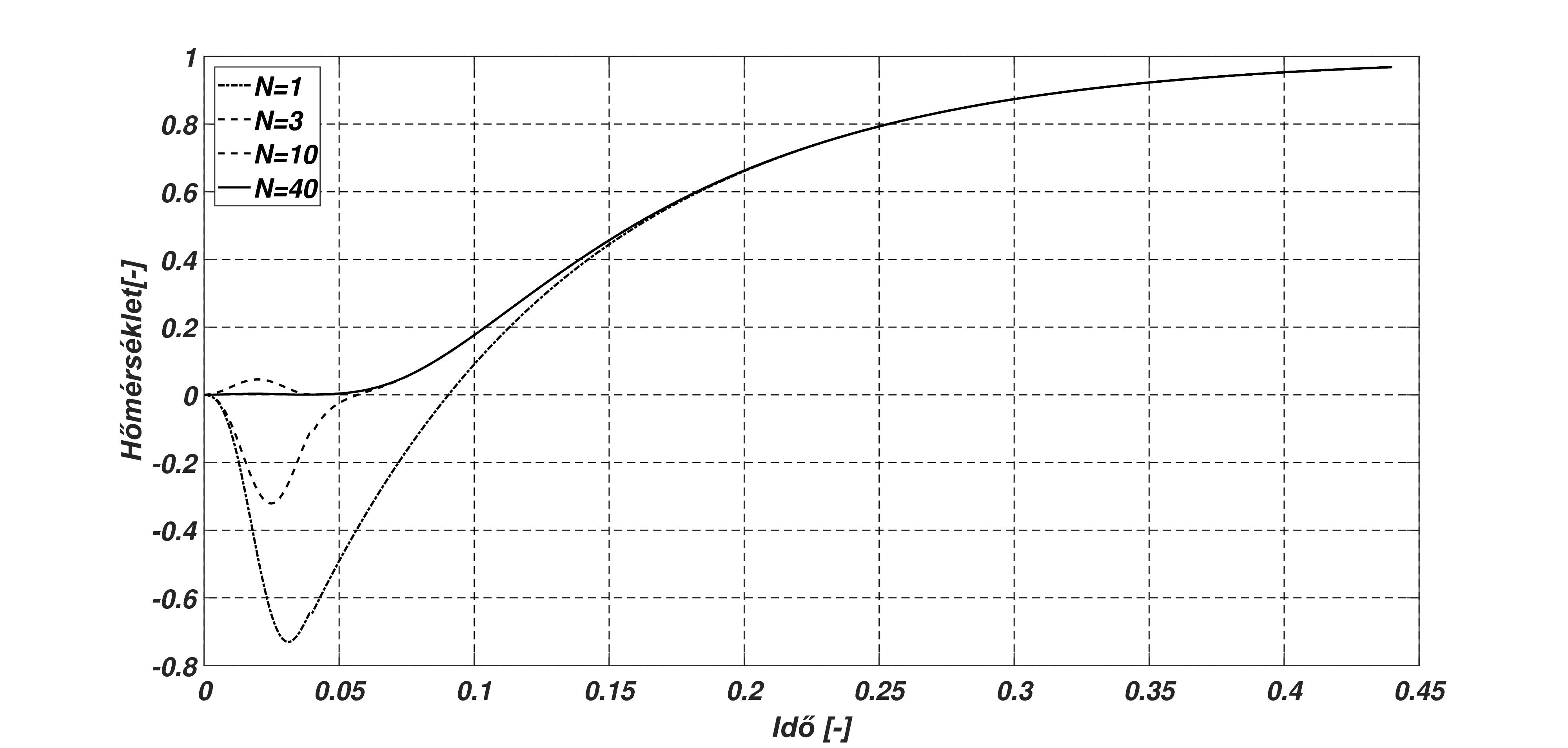}
\caption{The convergence of rear side temperature history considering more and more terms.}
\label{fig:analgk1}
\end{figure}

Since the rear side temperature history has importance during the evaluation of heat pulse experiments, let us check its convergence considering more and more terms in the sum (see Fig. \ref{fig:analgk1}). In this case the solution of Fourier equation is presented ($\tau_q = \kappa^2, \tau_\Delta=0.04$) and $N=1, 3, 10, 40$ terms are considered. It is visible that the initial region is considerably sensitive but the difference disappears after a certain time and the first term alone seems to be enough. 

\section{Validation of solution}
\begin{figure}
\includegraphics[width=12cm,height=7cm]{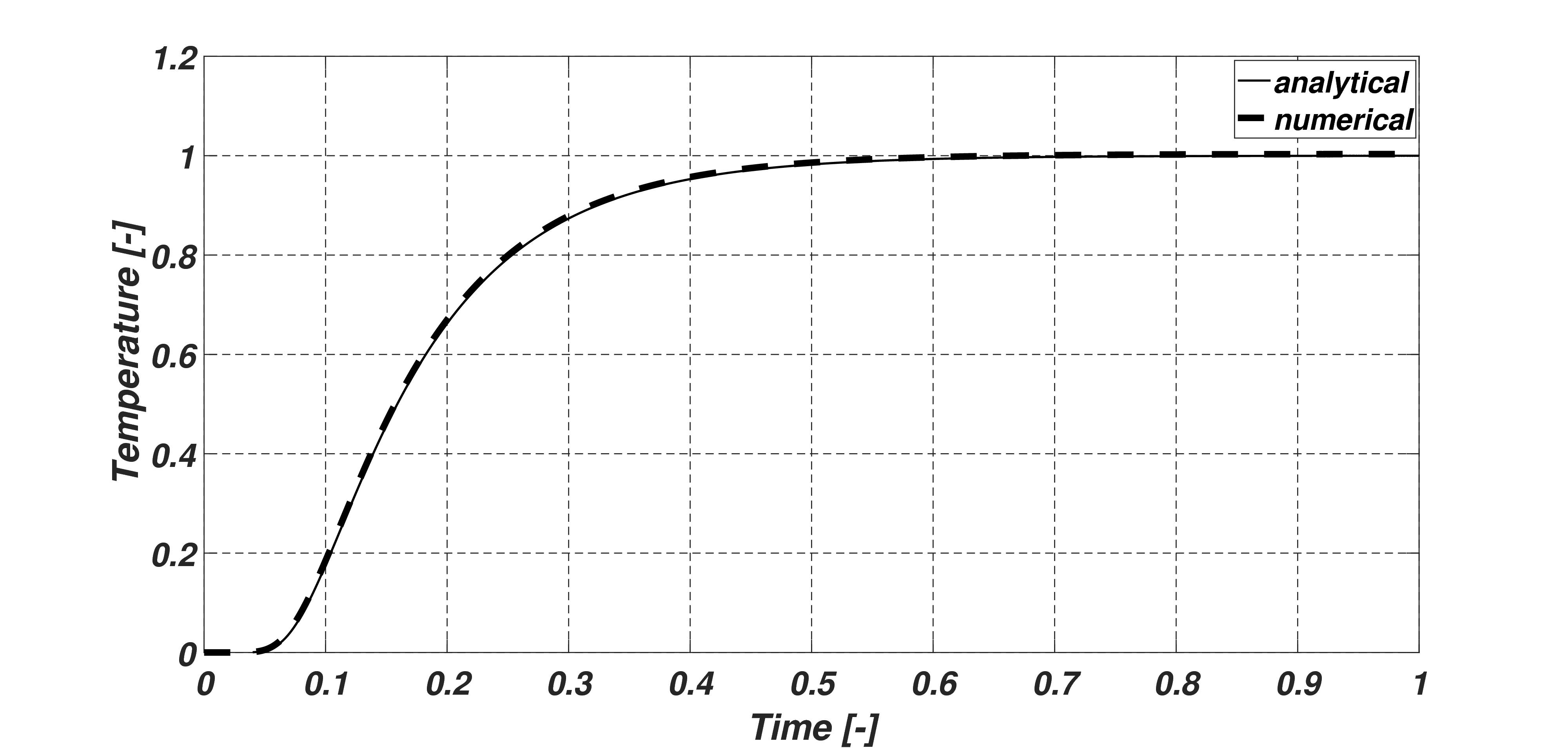}
\caption{The rear side temperature history considering $\tau_q=\kappa^2=0.02$, using $40$ terms. }
\label{fig:analgk2}
\end{figure}

\begin{figure}
\includegraphics[width=12cm,height=7cm]{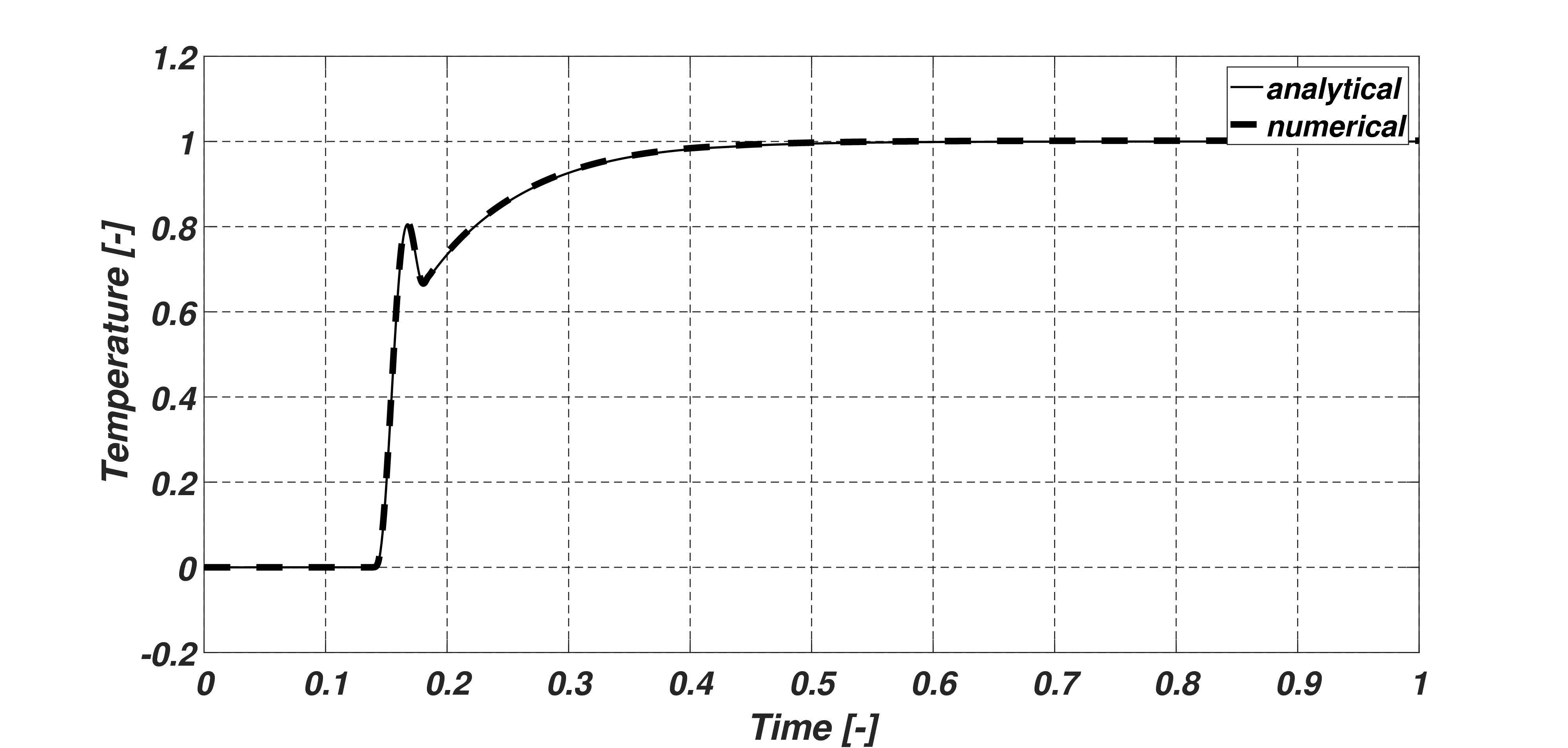}
\caption{The rear side temperature history considering $\tau_q=0.02$, $\kappa^2=0$, using $200$ terms. }
\label{fig:analgk3}
\end{figure}

\begin{figure}
\includegraphics[width=12cm,height=7cm]{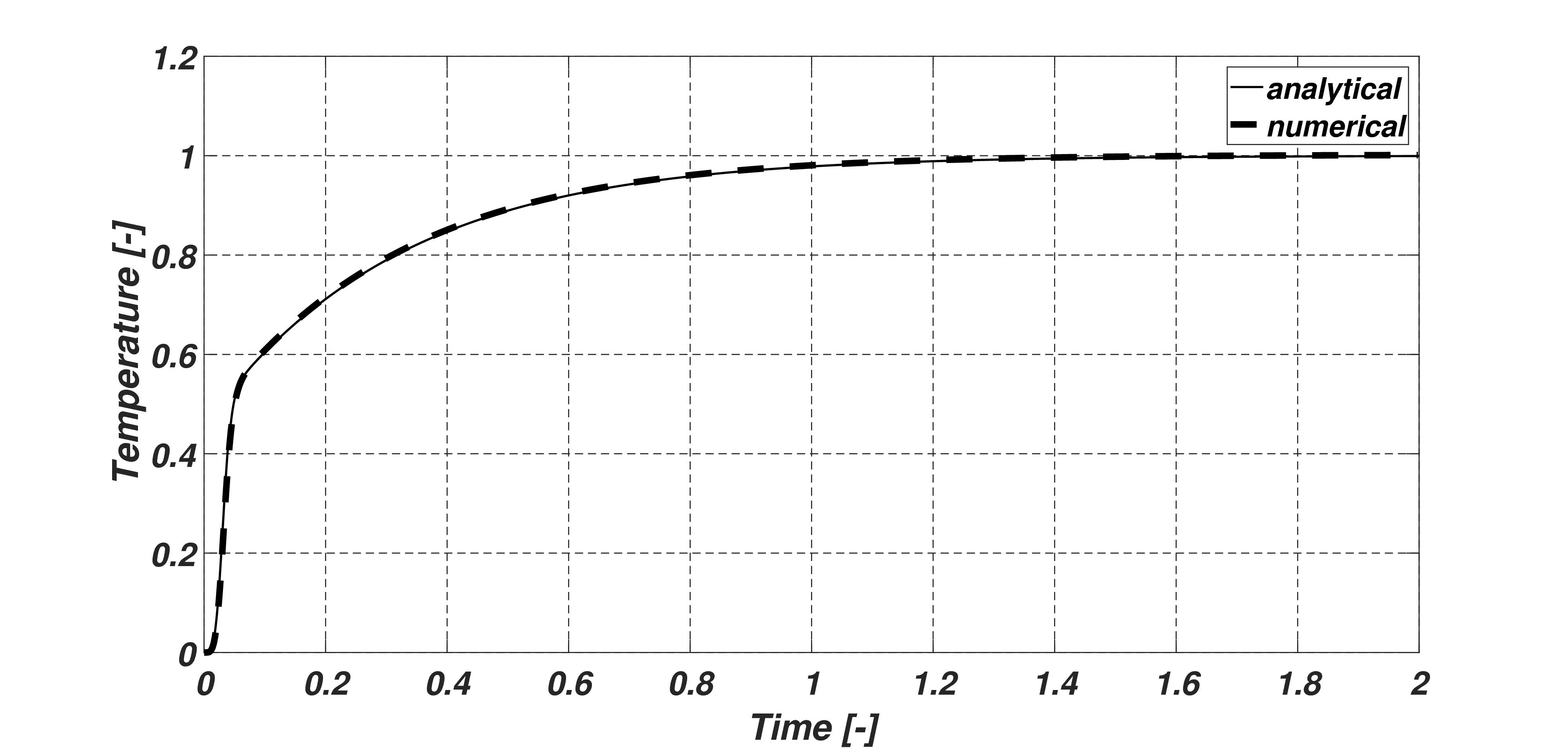}
\caption{The rear side temperature history considering $\tau_q=0.02$, $\kappa^2=0.2$, using $10$ terms. }
\label{fig:analgk4}
\end{figure}

The presented analytic solution is compared to the available numerical code \cite{KovVan15} as a validation (see Figs. \ref{fig:analgk2}, \ref{fig:analgk3} and \ref{fig:analgk4}). Naturally, the analytic solution runs much faster especially in the over-damped region ($\kappa^2>\tau_q$) without resulting in any unphysical temperature history. The over-damped solutions have greater importance as all the measurements confirm such behavior \cite{Botetal16, Vanetal17}.
The comparison is performed in three different  cases:
\begin{enumerate}
\item Fourier's solution: $\tau_q=\kappa^2=0.02$ (Fig. \ref{fig:analgk2}),
\item MCV's solution: $\tau_q=0.02$, $\kappa^2=0$ (Fig. \ref{fig:analgk3}),
\item Over-damped solution: $\tau_q=0.02$, $\kappa^2=0.2$ (Fig. \ref{fig:analgk4}). 
\end{enumerate}
The dimensionless pulse length is $\tau_\Delta=0.04$ in every case.

\section{Conclusions}
The analytic solution for Guyer-Krumhansl equation is presented considering finite heat pulse length on the front side and adiabatic condition on the rear side.  It should be emphasized that finite spatial region is also considered which makes the results more applicable for practical cases. The solution is obtained in the form of an infinite sum. It converges quickly to the exact solution in case of a smooth temperature history. In case of MCV equation, $200$ terms are sufficient to model the sharp wavefront.

It is easier to define boundary conditions for the field of heat flux and calculate the temperature field as a consequence. Applying the same idea for numerical codes leads to the shifted field concept described in \cite{KovVan15} and tested in several cases \cite{Botetal16, Vanetal17,  KovVan18dpl, KovVan18, KovVan16}. The analytical solution is validated by an explicit numerical method for every possible domain could appear in GK equation. Then the obtained analytical solution could be of a good use to investigate the entropy production paradox discussed by Barletta and Zanchini \cite{BarZan97a} in connection with the Taitel's paradox \cite{Taitel72}.

It was highlighted by Zhukovsky \cite{Zhukov16} that GK equation could violate the maximum principle under over-damped (or over-diffusive) conditions. Here, in the presented solutions the negative temperature domain does not exist even for the over-damped region. 

Now, one has to move on the more difficult case containing cooling boundary condition to widen possibilities.



\section{Acknowledgements}
\label{ackn}

The work was supported by the grant National Research, Development and Innovation Office – NKFIH, NKFIH 124366 and
NKFIH 124508.

\bibliographystyle{elsarticle-num} 

\begin{thebibliography}{10}
\expandafter\ifx\csname url\endcsname\relax
  \def\url#1{\texttt{#1}}\fi
\expandafter\ifx\csname urlprefix\endcsname\relax\def\urlprefix{URL }\fi
\expandafter\ifx\csname href\endcsname\relax
  \def\href#1#2{#2} \def\path#1{#1}\fi

\bibitem{Botetal16}
S.~Both, B.~Cz{\'e}l, T.~F{\"u}l{\"o}p, G.~Gr{\'o}f, {\'A}.~Gyenis,
  R.~Kov{\'a}cs, P.~V{\'a}n, J.~Verh{\'a}s, Deviation from the {F}ourier law in
  room-temperature heat pulse experiments, Journal of Non-Equilibrium
  Thermodynamics 41~(1) (2016) 41--48.

\bibitem{Vanetal17}
P.~V{\'a}n, A.~Berezovski, T.~F{\"u}l{\"o}p, G.~Gr{\'o}f, R.~Kov{\'a}cs,
  {\'A}.~Lovas, J.~Verh{\'a}s, Guyer-{K}rumhansl-type heat conduction at room
  temperature, EPL 118~(5) (2017) 50005, arXiv:1704.00341v1.

\bibitem{KovVan15}
R.~Kovács, P.~Ván, Generalized heat conduction in heat pulse experiments,
  International Journal of Heat and Mass Transfer 83 (2015) 613 -- 620.

\bibitem{Max1867}
J.~C. Maxwell, On the dynamical theory of gases, Philosophical Transactions of
  the Royal Society of London 157 (1867) 49--88.

\bibitem{Cattaneo58}
C.~Cattaneo, Sur une forme de lequation de la chaleur eliminant le paradoxe
  dune propagation instantanee, Comptes Rendus Hebdomadaires Des Seances De
  L'Academie Des Sciences 247~(4) (1958) 431--433.

\bibitem{Vernotte58}
P.~Vernotte, Les paradoxes de la th{\'e}orie continue de l{\'e}quation de la
  chaleur, Comptes Rendus Hebdomadaires Des Seances De L'Academie Des Sciences
  246~(22) (1958) 3154--3155.

\bibitem{JosPre89}
D.~D. Joseph, L.~Preziosi, Heat waves, Reviews of Modern Physics 61~(1) (1989)
  41.

\bibitem{JosPre90a}
D.~D. Joseph, L.~Preziosi, Addendum to the paper on heat waves, Reviews of
  Modern Physics 62~(2) (1990) 375--391.

\bibitem{Gyar77a}
I.~Gyarmati, On the wave approach of thermodynamics and some problems of
  non-linear theories, Journal of Non-Equilibrium Thermodynamics 2 (1977)
  233--260.

\bibitem{JouVasLeb88ext}
D.~Jou, J.~Casas-V{\'a}zquez, G.~Lebon, Extended irreversible thermodynamics,
  Reports on Progress in Physics 51~(8) (1988) 1105.

\bibitem{Tzou95}
D.~Y. Tzou, A unified field approach for heat conduction from macro- to
  micro-scales, Journal of Heat Transfer 117~(1) (1995) 8--16.

\bibitem{MulRug98}
I.~Müller, T.~Ruggeri, Rational Extended Thermodynamics, Springer, 1998.

\bibitem{VanFul12}
P.~Ván, T.~Fülöp, Universality in heat conduction theory -- weakly nonlocal
  thermodynamics, Annalen der Physik (Berlin) 524~(8) (2012) 470--478.

\bibitem{BerVan15}
A.~Berezovski, P.~Ván, Microinertia and internal variables, arXiv preprint
  arXiv:1504.03485.

\bibitem{Cimmelli09nl}
V.~Cimmelli, A.~Sellitto, D.~Jou, Nonlocal effects and second sound in a
  non-equilibrium steady state, Physical Review B 79~(1) (2009) 014303.

\bibitem{Cimm09diff}
V.~A. Cimmelli, Different thermodynamic theories and different heat conduction
  laws, Journal of Non-Equilibrium Thermodynamics 34~(4) (2009) 299--333.

\bibitem{Tisza38}
L.~Tisza, Transport phenomena in {H}elium {II}, Nature 141 (1938) 913.

\bibitem{Lan47}
L.~Landau, On the theory of superfluidity of {H}elium {II}, Journal of Physics
  11~(1) (1947) 91--92.

\bibitem{Pesh44}
V.~Peshkov, Second sound in {H}elium {II}, J. Phys. (Moscow) 381~(8).

\bibitem{GK64}
R.~A. Guyer, J.~A. Krumhansl, Dispersion relation for second sound in solids,
  Physical Review 133~(5A) (1964) A1411.

\bibitem{GuyKru66a1}
R.~A. Guyer, J.~A. Krumhansl, Solution of the linearized phonon {B}oltzmann
  equation, Physical Review 148~(2) (1966) 766--778.

\bibitem{GuyKru66a2}
R.~A. Guyer, J.~A. Krumhansl, Thermal conductivity, second sound and phonon
  hydrodynamic phenomena in nonmetallic crystals, Physical Review 148~(2)
  (1966) 778--788.

\bibitem{Van01a}
P.~Ván, Weakly nonlocal irreversible thermodynamics -- the {G}uyer-{K}rumhansl
  and the {C}ahn-{H}illiard equations, Physic Letters A 290~(1-2) (2001)
  88--92.

\bibitem{VanKovFul15}
T.~Fülöp, R.~Kov\'acs, P.~V\'an, Thermodynamic hierarchies of evolution
  equations, Proceedings of the Estonian Academy of Sciences 64~(3) (2015)
  389--395.

\bibitem{KovVan18dpl}
R.~Kov{\'a}cs, P.~V{\'a}n, Thermodynamical consistency of the {D}ual {P}hase
  {L}ag heat conduction equation, Continuum Mechanics and
  ThermodynamicsManuscript.

\bibitem{JacWalMcN70}
H.~E. Jackson, C.~T. Walker, T.~F. McNelly, Second sound in {N}a{F}, Physical
  Review Letters 25~(1) (1970) 26--28.

\bibitem{JacWal71}
H.~E. Jackson, C.~T. Walker, Thermal conductivity, second sound and
  phonon-phonon interactions in {N}a{F}, Physical Review B 3~(4) (1971)
  1428--1439.

\bibitem{McNEta70a}
T.~F. McNelly, S.~J. Rogers, D.~J. Channin, R.~J. Rollefson, W.~M. Goubau,
  G.~E. Schmidt, J.~A. Krumhansl, R.~O. Pohl, Heat pulses in {N}a{F}: onset of
  second sound, Physical Review Letters 24~(3) (1970) 100--102.

\bibitem{McN74t}
T.~F. McNelly, Second {S}ound and {A}nharmonic {P}rocesses in {I}sotopically
  {P}ure {A}lkali-{H}alidesPh.D. Thesis, Cornell University.

\bibitem{DreStr93a}
W.~Dreyer, H.~Struchtrup, Heat pulse experiments revisited, Continuum Mechanics
  and Thermodynamics 5 (1993) 3--50.

\bibitem{Ma13a}
Y.~Ma, A transient ballistic–diffusive heat conduction model for heat pulse
  propagation in nonmetallic crystals, International Journal of Heat and Mass
  Transfer 66 (2013) 592--602.

\bibitem{Ma13a1}
Y.~Ma, A {H}ybrid {P}honon {G}as {M}odel for {T}ransient
  {B}allistic-{D}iffusive {H}eat {T}ransport, Journal of Heat Transfer 135~(4)
  (2013) 044501.

\bibitem{Ma13a2}
Y.~Ma, Equation of phonon hydrodynamics for non-{F}ourier heat conduction, in:
  44th AIAA Thermophysics Conference, 2013, p. 2902.

\bibitem{KovVan18}
R.~Kov{\'a}cs, P.~V{\'a}n, Second sound and ballistic heat conduction: {N}a{F}
  experiments revisited, International Journal of Heat and Mass Transfer 117
  (2018) 682--690, submitted, arXiv preprint arXiv:1708.09770.

\bibitem{Nyiri91}
B.~Nyíri, On the entropy current, Journal of Non-Equilibrium Thermodynamics
  16~(2) (1991) 179--186.

\bibitem{MitEta95}
K.~Mitra, S.~Kumar, A.~Vedevarz, M.~K. Moallemi, Experimental evidence of
  hyperbolic heat conduction in processed meat, Journal of Heat Transfer
  117~(3) (1995) 568--573.

\bibitem{TilVic09}
E.~P. Scott, M.~Tilahun, B.~Vick, The question of thermal waves in
  heterogeneous and biological materials, Journal of Biomechanical Engineering
  131~(7) (2009) 074518.

\bibitem{HerBec00}
H.~Herwig, K.~Beckert, Fourier versus non-{F}ourier heat conduction in
  materials with a nonhomogeneous inner structure, Transactions-American
  Society of Mechanical Engineers Journal of Heat Transfer 122~(2) (2000)
  363--364.

\bibitem{HerBec00b}
H.~Herwig, K.~Beckert, Experimental evidence about the controversy concerning
  {F}ourier or non-{F}ourier heat conduction in materials with a nonhomogeneous
  inner structure, Heat and Mass Transfer 36~(5) (2000) 387--392.

\bibitem{TanEtal07}
D.~Tang, N.~Araki, N.~Yamagishi, Transient temperature responses in biological
  materials under pulsed {IR} irradiation, Heat and Mass Transfer 43~(6) (2007)
  579--585.

\bibitem{AkbPas14}
A.~H. Akbarzadeh, D.~Pasini, Phase-lag heat conduction in multilayered cellular
  media with imperfect bonds, International Journal of Heat and Mass Transfer
  75 (2014) 656--667.

\bibitem{AfrinEtal12}
N.~Afrin, J.~Zhou, Y.~Zhang, D.~Y. Tzou, J.~K. Chen, Numerical simulation of
  thermal damage to living biological tissues induced by laser irradiation
  based on a generalized dual phase lag model, Numerical Heat Transfer, Part A:
  Applications 61~(7) (2012) 483--501.

\bibitem{LiuChen10}
K.-C. Liu, H.-T. Chen, Investigation for the dual phase lag behavior of
  bio-heat transfer, International Journal of Thermal Sciences 49~(7) (2010)
  1138--1146.

\bibitem{Zhang09}
Y.~Zhang, Generalized dual-phase lag bioheat equations based on nonequilibrium
  heat transfer in living biological tissues, International Journal of Heat and
  Mass Transfer 52~(21) (2009) 4829--4834.

\bibitem{Ruk14}
S.~A. Rukolaine, Unphysical effects of the dual-phase-lag model of heat
  conduction, International Journal of Heat and Mass Transfer 78 (2014) 58--63.

\bibitem{Ruk17}
S.~A. Rukolaine, Unphysical effects of the dual-phase-lag model of heat
  conduction: higher-order approximations, International Journal of Thermal
  Sciences 113 (2017) 83--88.

\bibitem{Fabetal14}
M.~Fabrizio, F.~Franchi, Delayed thermal models: stability and thermodynamics,
  Journal of Thermal Stresses 37~(2) (2014) 160--173.

\bibitem{FabLaz14a}
M.~Fabrizio, B.~Lazzari, Stability and second law of thermodynamics in
  dual-phase-lag heat conduction, International Journal of Heat and Mass
  Transfer 74 (2014) 484--489.

\bibitem{FabEtal16}
M.~Fabrizio, B.~Lazzari, V.~Tibullo, Stability and thermodynamic restrictions
  for a dual-phase-lag thermal model, Journal of Non-Equilibrium
  ThermodynamicsPublished Online:2017/01/10.

\bibitem{Quin07}
R.~Quintanilla, R.~Racke, Qualitative aspects in dual-phase-lag heat
  conduction, Proceedings of the Royal Society of London A: Mathematical,
  Physical and Engineering Sciences 463~(2079) (2007) 659--674.

\bibitem{ChirCiaTib17}
S.~Chiriţă, M.~Ciarletta, V.~Tibullo, On the thermomechanical consistency of
  the time differential dual-phase-lag models of heat conduction, International
  Journal of Heat and Mass Transfer 114 (2017) 277 -- 285.

\bibitem{Zhukov16}
K.~Zhukovsky, Violation of the maximum principle and negative solutions for
  pulse propagation in {G}uyer--{K}rumhansl model, International Journal of
  Heat and Mass Transfer 98 (2016) 523--529.

\bibitem{Zhu16a}
K.~V. Zhukovsky, Exact solution of {G}uyer--{K}rumhansl type heat equation by
  operational method, International Journal of Heat and Mass Transfer 96 (2016)
  132--144.

\bibitem{Zhu16b}
K.~V. Zhukovsky, Operational approach and solutions of hyperbolic heat
  conduction equations, Axioms 5~(4) (2016) 28.

\bibitem{ZhuSri17}
K.~V. Zhukovsky, H.~M. Srivastava, Analytical solutions for heat diffusion
  beyond {F}ourier law, Applied Mathematics and Computation 293 (2017)
  423--437.

\bibitem{GrofPhD02}
G.~I. Gr{\'o}f, Homog{\'e}n {\'e}s k{\'e}tr{\'e}teg{\H{u}} mint{\'a}k
  h{\H{o}}m{\'e}rs{\'e}kletvezet{\'e}si t{\'e}nyez{\H{o}}j{\'e}nek
  m{\'e}r{\'e}se flash m{\'o}dszerrel.

\bibitem{BCTFGGPV13}
B.~Cz\'el, T.~Fülöp, G.~Gr\'of, P.~V\'an, Comparison of temperature responses
  of the laser flash method in case of parabolic and hyperbolic heat conduction
  models, in: D.~Sz. (Ed.), 11th International Conference on Heat Engines and
  Environmental Protection, Balatonfüred, BME, Dep. of Energy Engineering,
  Budapest, 2013, pp. 133--139.

\bibitem{CarJae59b}
H.~S. Carslaw, J.~C. Jaeger, Conduction of heat in solids, Oxford: Clarendon
  Press, 1959, 2nd ed.

\bibitem{Farlow93b}
S.~J. Farlow, Partial differential equations for scientists and engineers,
  Courier Corporation, 1993.

\bibitem{GTvN11b}
G.~T. von Nessi, Analytic {M}ethods in {P}artial {D}ifferential {E}quations.

\bibitem{KovVan16}
R.~Kovács, P.~Ván, {M}odels of {B}allistic {P}ropagation of {H}eat at {L}ow
  {T}emperatures, International Journal of Thermophysics 37~(9) (2016) 95.

\bibitem{BarZan97a}
A.~Barletta, E.~Zanchini, Hyperbolic heat conduction and local equilibrium: a
  second law analysis, International Journal of Heat and Mass Transfer 40~(5)
  (1997) 1007--1016.

\bibitem{Taitel72}
Y.~Taitel, On the parabolic, hyperbolic and discrete formulation of the heat
  conduction equation, Internation Journal of Heat and Mass Transfer 15 (1972)
  369--371.

\end{thebibliography}





\end{document}